\documentclass{iaus}
\usepackage{graphicx}
\usepackage{color}

\title[The Evolution of a Double Diffusive Magnetic Buoyancy Instability] %% give here short title %%
{The Evolution of a Double Diffusive Magnetic Buoyancy Instability}

\author[Silvers, Vasil, Brummell \& Proctor]   %% give here short author list %%
{Lara J. Silvers$^1$, Geoffrey M. Vasil$^2$, Nicholas H. Brummell$^3$ \and Michael R. E. Proctor$^4$}

\affiliation{$^1$ Centre for Mathematical Science, City University London, Northampton Square,
London, EC1V 0HB, U. K.   \\ email: {\tt lara.silvers.1@city.ac.uk} \\[\affilskip]
$^2$Canadian Institute for Theoretical Astrophysics, 60 St. George Street, Toronto, ON M5S 3H8, Canada \\email: {\tt vasil@cita.utoronto.ca}
 \\[\affilskip]
$^3$Department of Applied Mathematics \& Statistics, University of California, Santa Cruz, CA 95064, U.S.A. \\email: {\tt brummell@soe.ucsc.edu}
 \\[\affilskip]
$^4$Department of Applied Mathematics and Theoretical Physics, University of Cambridge, Wilberforce Road, Cambridge, CB3 0WA, U. K. \\email: {\tt mrep@damtp.cam.ac.uk}
}

\pubyear{2010 }
\volume{271}  %% insert here IAU Symposium No.
\pagerange{111--111}
\jname{Astrophysical Dynamics: From Stars to Galaxies.}
\editors{N. H. Brummell \& A. S. Brun, eds.}
\begin{document}

\maketitle

\begin{abstract}
Recently, \cite{silversetal}, using numerical simulations, confirmed the existence of a double diffusive magnetic buoyancy instability of a layer of horizontal magnetic field produced by the interaction of a shear velocity field with a weak vertical field.  Here, we demonstrate the longer term nonlinear evolution of such an instability in the simulations.  We find that a quasi two-dimensional interchange instability rides (or ``surfs'') on the growing shear-induced background downstream field gradients.  The region of activity expands since three-dimensional perturbations remain unstable in the wake of this upward-moving activity front, and so the three-dimensional nature becomes more noticeable with time.
\end{abstract}

\firstsection % if your document starts with a section,
\section{Introduction}

Although at this time there is no general consensus on all the elements necessary for the operation of a solar dynamo that creates the large-scale field that we observe as solar active regions, one ingredient that is generally considered to play an important role in all the proposed scenarios (see e.g. \cite{parker_1993,babcock_1961,Leighton_1969}) is magnetic buoyancy.    It is widely believed that the strong shear in the solar tachocline can stretch existing poloidal magnetic field into strong toroidal magnetic field, and then magnetic buoyancy instabilities of the toroidal field create compact toroidal magnetic structures that rise into the overlaying convection zone to either be re-processed there or to emerge at the solar surface as active regions.

It is thus of considerable interest to seek to understand the formation and evolution of such magnetic structures. Until fairly recently, such investigations  have principally focussed on the evolution of pre-conceived, idealised buoyant structures (e.g.\  \cite{parker_1955}, \cite{fan_etal_1998}, \cite{emonet_moreno-insertis_1998}, \cite{hughes_etal_1998}, \cite{hughes_falle_1998}, \cite{wissink_etal_2000}, \cite{fan_etal_2003}, \cite{abbett_etal_2004}, \cite{JouveBrun}).
Recently attention turned to the problem of the self-consistent
generation of such structures as well as their evolution, in
particular using the dynamics that are expected to be available in the
solar tachocline, i.e.\ strong velocity shear
(\cite{brummell_etal_2002}, \cite{cline_etal_2003a},
\cite{cline_etal_2003b}, \cite{cattaneo_etal_2006}, \cite{Vasil1},
\cite{Vasil2}, \cite{silversetal2}, \cite{silversetal}).  For different configurations of shear and seed poloidal magnetic field, the creation of buoyantly-unstable magnetic layers that generate rising compact magnetic structures can be possible.

However, as was demonstrated in \cite{Vasil1} and \cite{Vasil2},
for the case of an initial vertical (radial) poloidal field, the mere
existence of a strong velocity shear does not guarantee the formation
of a buoyant structures, as is often assumed in discussions of the
large-scale solar dynamo. The instability will only occur for certain
values of the  parameters of the problem.  Of particular importance
are the strength and geometry of the shear flow as related to the
background stratification via a Richardson number.   In \cite{Vasil1},
it was shown that sufficiently strong shear forcing could lead to the
generation of buoyant magnetic structures. However, the forcings
required were extreme and unphysical for the context of the solar tachocline.   Whilst attempting an analytical understanding of this
problem, \cite{Vasil2} outlined the possibility that the stabilizing effect of stratification could be mitigated by a strong difference between the
stabilizing thermal diffusion time and the destabilizing magnetic
diffusion time, as had been noted earlier \cite{hughes_1985}  for the simpler problem of the instability of a steady stratified horizontal magnetic field.   When the ratio of these diffusivities $\zeta=\eta /
\kappa$ (where $\eta$ is the magnetic diffusivity, and $\kappa$ is the
thermal diffusivity) is small, instability may occur for more
reasonable shear rates, via a double-diffusive
instability mechanism.

\begin{figure*}
\includegraphics[width={\linewidth},height={\textheight}]{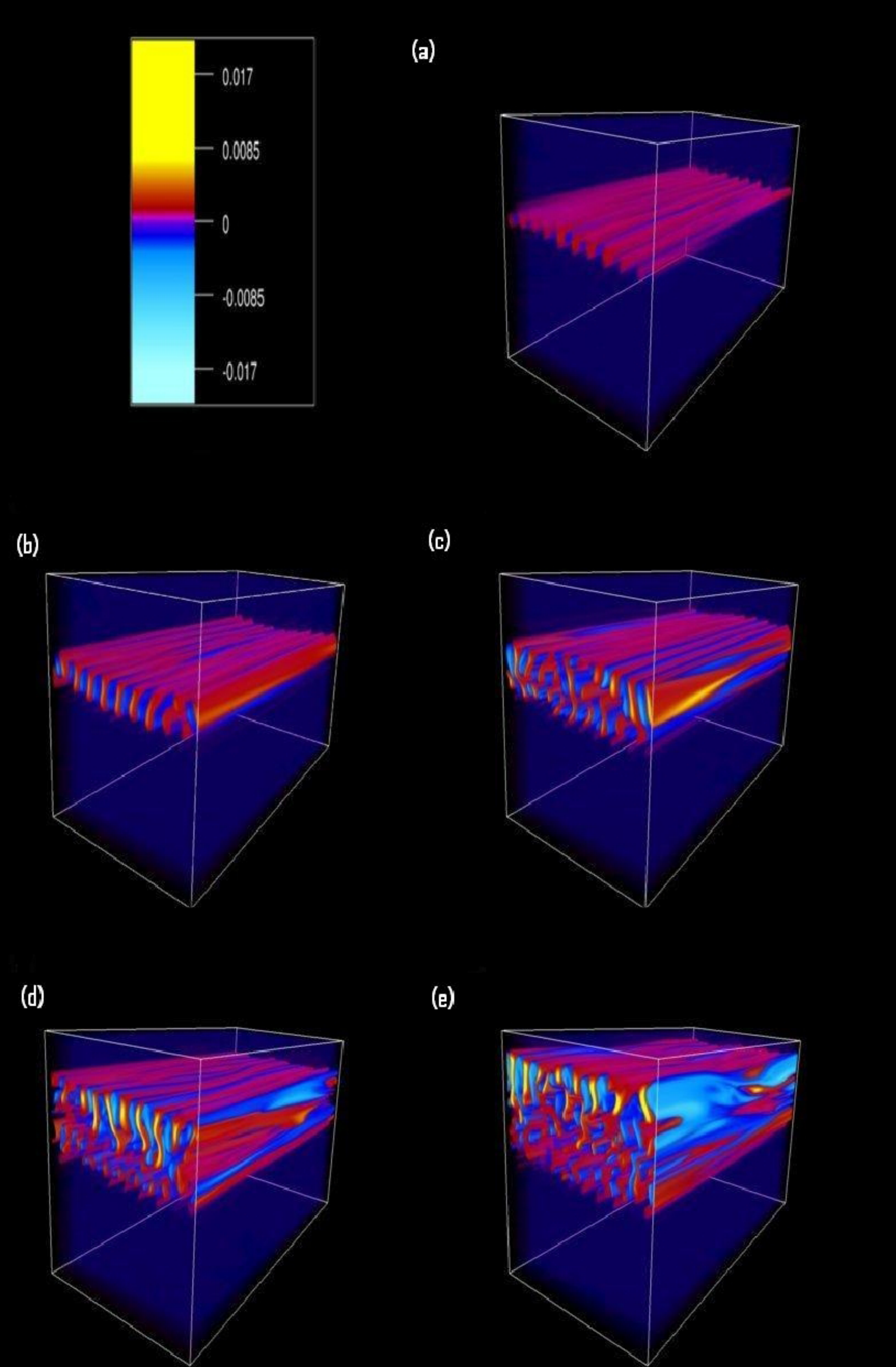}
\caption{Volume rendered images of the vertical component of velocity at: (a)~$t=112.73$; (b)~$t=135.63$; (c)~$t=158.53$; (d)~$t=181.43$; (e)~$t=204.33$. }
\label{figure1}
\end{figure*}

This possibility was explored using numerical simulations in \cite{silversetal}.  This paper provided clear evidence for the existence of the double diffusive magnetic buoyancy in the configuration of \cite{Vasil1} for physically-realizable forcings, and showed that the onset of this instability is quasi two-dimensional in nature, in agreement with the previous linear stability calculations of \cite{tobiashughes}.   However, the purpose of \cite{silversetal} was to establish the existence of the instability.  Here, we wish to examine the subsequent non-linear evolution of such an instability. After substantial further computation, we present the results of the evolution of Case~1 from \cite{silversetal} until the point where interactions with the boundaries of the computational domain invalidate further exploration.

\section{Mathematical Modeling}

The basic set up for this model is similar to that of several previous
studies of magnetic buoyancy in a fully compressible plane layer
(e.g.\ \cite{Vasil1}, \cite{silversetal2}) and exactly the same as that described in
\cite{silversetal}. The standard equations for compressible fluid
dynamics in dimensionless form (see e.g.\  \cite{MPW}) are augmented
with a forcing term that ensures the maintenance of a desired target
shear flow against viscous decay in the absence of magnetic effects.
The chosen target shear flow here (as in \cite{silversetal}) is of the form $\textbf{u}=(U_{z},0,0)$ with $U(z)$ a hyperbolic tangent function, selected to be representative of the smooth radial shear transition believed to exist in the solar tachocline.  Initially, we impose a weak uniform vertical magnetic field and examine the action of the shear on this seed field.

The problem is defined by a number of dimensionless quantities that
appear in the governing equations and here we once again set these to
be the values for Case 1 in \cite{silversetal}. That is, $\zeta=5.0
\times 10^{-4}$,  $\sigma=2.5 \times 10^{-4}$, $C_{K}=5 \times 10^{-4}$,
$\alpha=1.25 \times 10^{-5}$, $\theta=5.0$, $m=1.6$, with aspect ratio for the computational domain 2:1:1 so that the domain is elongated in the direction of the shear flow.
The computation is performed by standard pseudospectral methods and was run on over 200 cores of a machine composed of 2GHz dual-core Opteron 270 processors for a few months.

\section{Results}

In order to investigate the full nonlinear evolution of a double diffusive magnetic buoyancy instability we evolve Case 1 from \cite{silversetal} until there is significant interaction with the upper boundary and the computation has to be halted. This occurs at $t=220$ in our dimensionless units.

\begin{description}
  \item[Three-Dimensionality: ]  Figure \ref{figure1} shows volume renderings of the vertical component of velocity at a series of times that range from a short while after the onset of the instability to shortly before the run was terminated.    Existence of a vertical velocity shows evidence of a mode of motion that is not driven by the forcing but which must spontaneously arise as an instability.  As described in \cite{silversetal}, initially the instability appears as a wavelike perturbation with high frequency in the cross-stream $y$ direction, with some but little variation in the $x$ direction.  These quasi two-dimensional rolls of motion aligned in the downstream direction are suggestive of the fastest growing modes that might be expected from the linear stability of a simple plane layer of magnetic field in a velocity shear (\cite{tobiashughes}).  As time advances, the region of instability increases in size, extending upwards towards the top of the computational domain.  It appears as though the strongest motions are occurring towards the top of this expanding region, but that action continues behind even after the peak of activity has passed on.

  \begin{figure}
\begin{center}
\includegraphics[width=100mm,height=80mm]{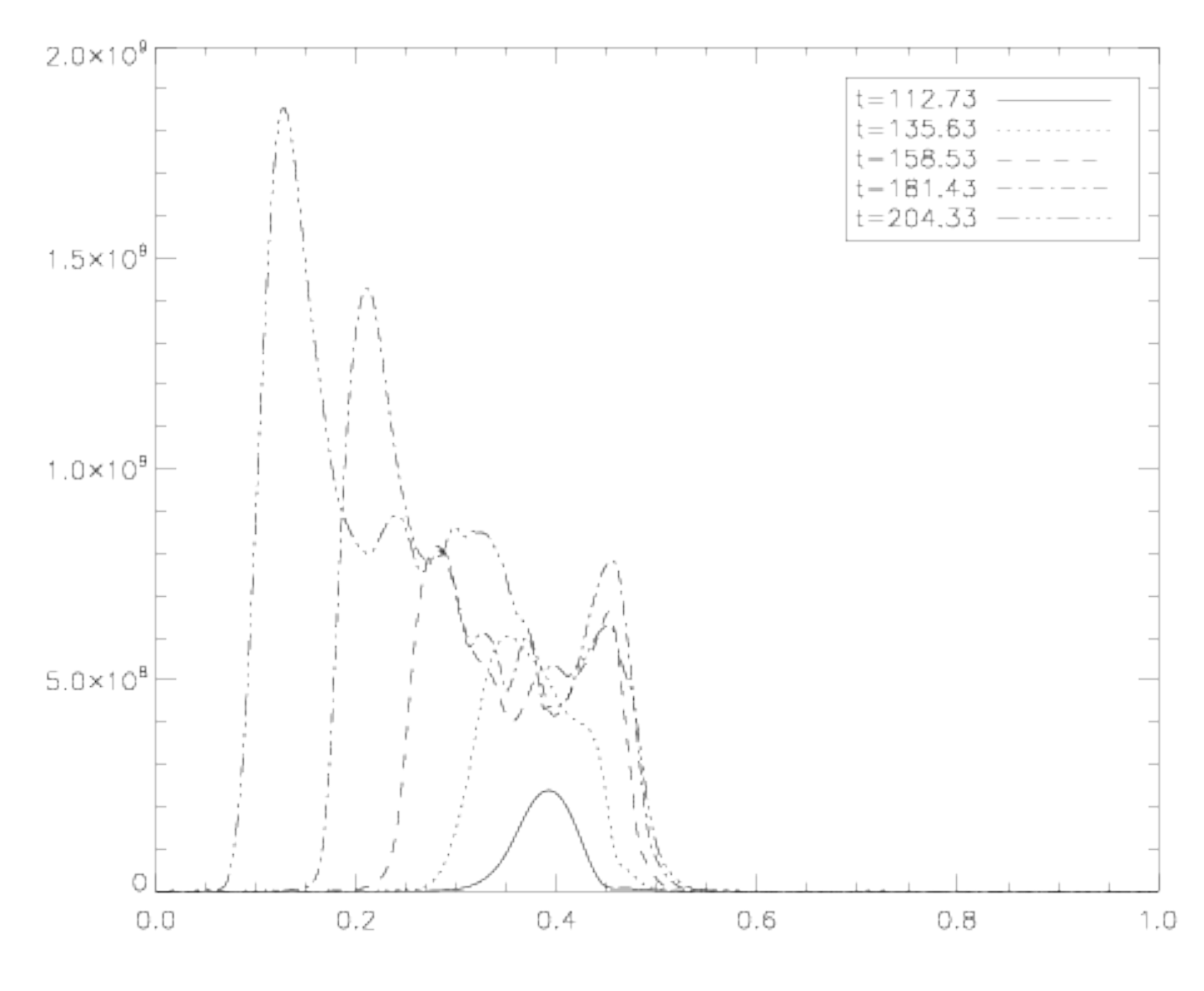}
\includegraphics[width=100mm,height=80mm]{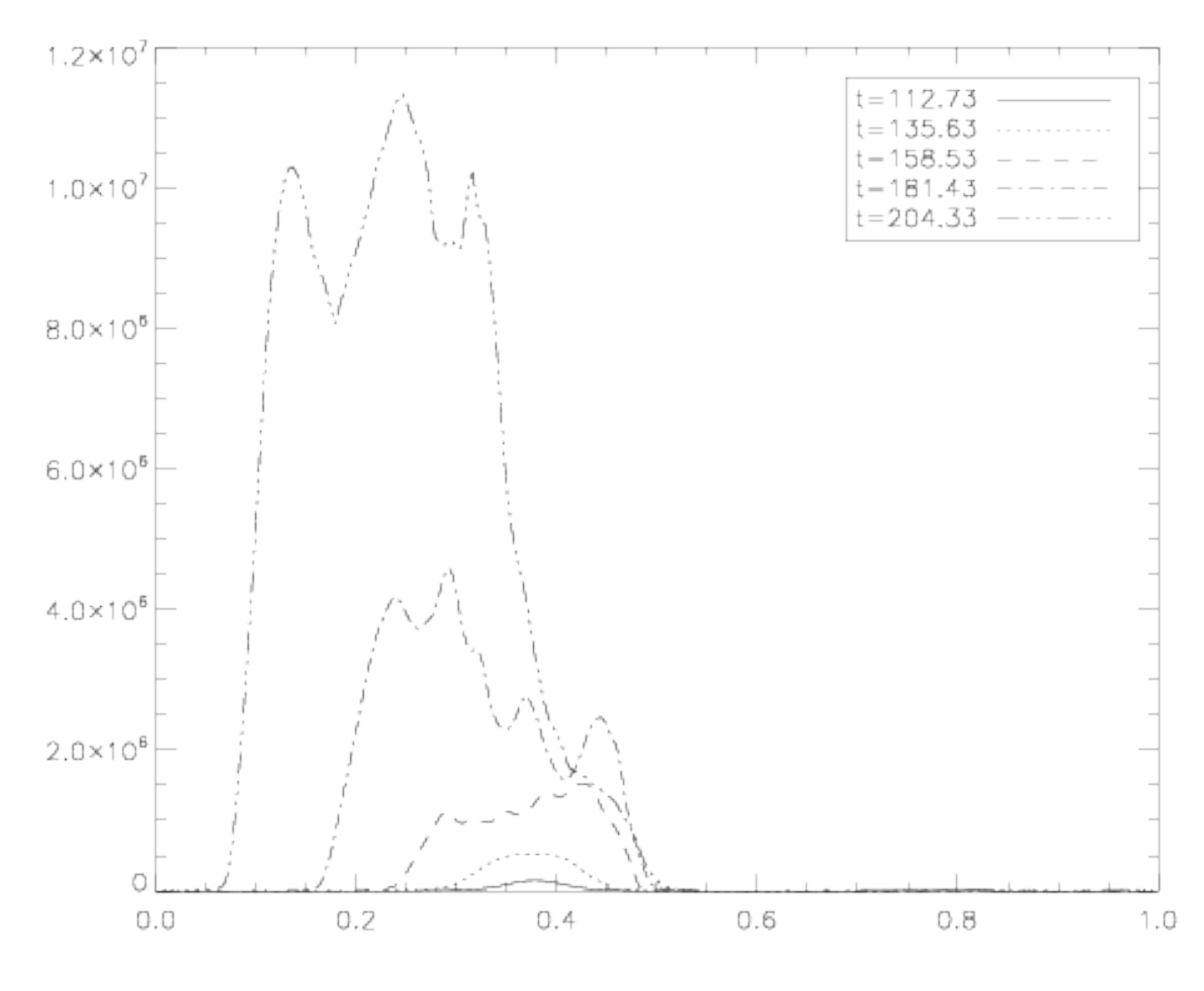}
\caption{({\sl a. Top panel}) $(\partial B_x/\partial y)^2/\alpha$ horizontally-averaged as a function of depth at various times, showing the evolution of two-dimensional disturbances.
({\sl b. Bottom panel})  $(\partial B_x/\partial x)^2/\alpha$ horizontally-averaged as a function of depth at various times, showing the evolution of three-dimensional disturbances.}
\label{figure2}
\end{center}
\end{figure}

 It also appears as though motions become more three dimensional as time progresses.  To check this out, we examine the quantities shown in Figures \ref{figure2}.  Initially, there is only a vertical magnetic field present. The action of the forced velocity profile is to draw out this initial field to induce a horizontal component in the region of shear.  Eventually the gradients induced are sufficient for the double-diffusion magnetic buoyancy instability to occur.  During the stretching phase, there are only two components of the magnetic field, the initial $B_z$ and the induced $B_x$.   When instability occurs, perturbations to all components of ${\bf B}$ may be present.  If the instability is two-dimensional, we expect to see a signature of the perturbation to $B_x$ in the $y$ direction with little perturbation in the $x$ direction.  If three-dimensional motions appear, we expect to see a signature of the perturbation to $B_x$ in the $x$ direction too.  We therefore examine $(\partial B_x/\partial y)^2$ and $(\partial B_x/\partial x)^2$ as measures of the existence of two- and three-dimensional perturbations.   From Figure \ref{figure2}a, clearly the strongest two-dimensional perturbation grow with time, and appear at the top of the expanding region of activity, as was seen visually in Figure \ref{figure1}.   Two-dimensional perturbations exist in the whole region, but are twice as strong near the top.   From Figure \ref{figure2}b, three-dimensional perturbations appear to exist more homogeneously throughout the activity region.  These perturbations exist for all times, but their measure is at least two orders of magnitude weaker than that of the two-dimensional perturbations.   However, the relative strength of the three-dimensional measure does appear to increase approximately four-fold over time, whilst remaining substantially weaker overall, confirming our visual inspection of Figure \ref{figure1}.

\begin{figure*}
\includegraphics[width=\linewidth,height=\textheight]{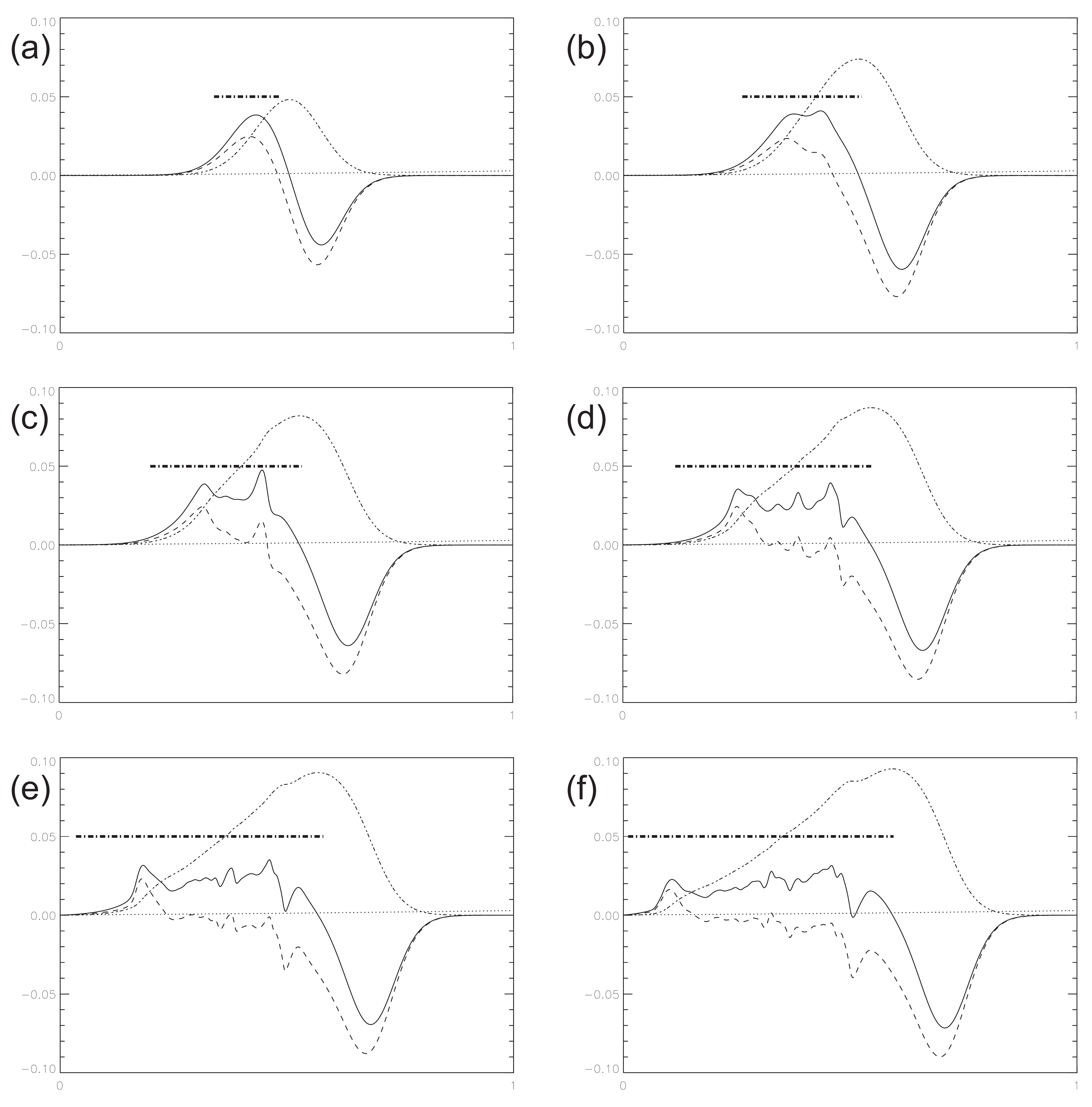}
\caption{Images showing the depth dependency of the horizontal
  averages of the following quantities:  $\alpha B_x\frac{dB_x}{dz}$(solid line),
  $\alpha \rho B_x\frac{d}{dz}\left(\frac{B_x}{\rho}\right)$(long-dashed line) and
  $-\zeta \frac{1}{\rho^{-\gamma}}\frac{d}{dz}\rho^{1-\gamma}T$(dotted
  line) at a series of times. The times for each panel are: (a)~$t=71.53$;(b)~$t=112.73$; (c)~$t=135.63$; (d)~$t=158.53$; (e)~$t=181.43$; (f)~$t=204.33$. These figures also show the region where $|w|$ (averaged in $x-y$ is greater than $1 \times 10^{-7}$ (dash-dot thick horizontal lines) where the instability could be considered to be occurring in the simulation, and the horizontal average of $5 \alpha B_x^2$ (short-dot-dashed line; scaling for clarity) to show the position of the growing background magnetic layer on which the instability ``surfs".}
\label{figure3}
\end{figure*}

  \item[The Depth of the Unstable Region: ]    The point where the maximum activity is obtained moves towards the upper boundary.  This appears to be a consequence of the underlying evolution of the induced $B_x$ field.  The shear induction of  $B_x$ produces an increasing peak of $B_x$ which leads to a rising ``front'' in the vertical gradient of $B_x$, on which the instability appears to ``surf''.  This evolving peak can be seen as the short dot-dashed line in panels of Figure~\ref{figure3}.  Instability continues behind this front but appears to be strongest where the front is passing through.  Eventually this front hits the upper boundary and at this point we can no longer continue the simulation.  The problem under consideration therefore has the character of a dynamic bifurcation, where the background state that determines the linear instability is evolving with time. The timescale for the development of the basic horizontal field structure here appears to be comparable with the growth rate of the $y$-dependent disturbances, somewhat complicating matters. Nonetheless it is to be expected that the disturbance will grow principally in regions where the background state is locally unstable.  From linear stability arguments (\cite{newcomb1961,tobiashughes}), we expect the fastest growing modes to have high cross-stream wavemnumber ($k_y$) and a downstream wavenumber ($k_x$) that is as long as possible -- quasi interchange modes.  We therefore here compare the extent of the instability found in our simulations with both the local two-dimensional interchange and three-dimensional criteria which would hold for a non-evolving stationary field configuration. These can be written, in a simplified form as

$$ \alpha B_x\frac{dB_x}{dz}>-\zeta
\frac{1}{\rho^{-\gamma}}\frac{d}{dz}\rho^{1-\gamma}T \hspace{0.4cm}
\textrm{(3D)}$$
$$ \alpha \rho
B_x\frac{d}{dz}\left(\frac{B_x}{\rho}\right)>-\zeta
\frac{1}{\rho^{-\gamma}}\frac{d}{dz}\rho^{1-\gamma}T \hspace{0.4cm}
\textrm{(interchange)} $$

\noindent (see e.g. \cite{hughes_2007} for further details). Note that the right hand side of both inequalities is the same. Figure \ref{figure3} shows the regions where these
inequalities are satisfied, namely where the lines corresponding to relevant field gradient criteria (left hand sides above) lie above the dotted line giving the stabilizing entropy gradient (right hand sides above). Initially, there is
considerable overlap between the regions were both the two and three
dimensional instability criteria are met, and as expected, the three-dimensional instability is slightly preferred. However, at later times we
see that the region where the two dimensional instability criterion is
met is confined to a very small region at the top of the expanding region. 
By contrast the region where the three
dimensional instability criterion is satisfied occupies almost half the
depth of the box at late times. The figure also shows the region where $|w|$(averaged in $x-y$) is significant (formally greater than $1 \times 10^{-7}$) to indicate the vertical extent of activity at each time. It can be seen that this expands reasonably in line with the extending region of (static) three-dimensional instability, and occupies at any time basically the whole region where there is significant positive gradient in $B_x^2$ since the stabilising entropy gradient is so low.

\end{description}

\section{Discussion}

From this simple analysis, one might hesitantly conclude that our initial visual impression is correct.  Initially, the instability does appear to be dominated by quasi two-dimensional modes, although formally both (stationary not dynamic) instability criteria (for interchange and three-dimensional modes) are satisfied.   The variation in $y$ of the instability is on a rather shorter scale than that appearing for larger values of the diffusivity ratio, as shown in the results of \cite{silversetal2}. This short transverse scale is reminiscent of the salt finger instability that occurs in the oceans when warm salty water overlies cold fresh water.  As the simulation progresses, the evolving region of the background field that provides the magnetic gradients to drive the instability dynamically creates a shifting and adjusting region of local instability.  The strongest instability occurs at the top of this region, and appears to still have a strong quasi two-dimensional component, whereas the motions in the wake of this front appear to be more likely dominated by a variety of unstable three-dimensional modes as the pure interchange modes become damped.  Overall, the region of instability broadens.

Considerably more work is required in order to understand fully the evolution of this instability.  Taller boxes or domains that adjust to allow the rising front of activity are required for a complete picture of the long time evolution and saturation of this process.  The efficiency of this process in its transport of magnetic field is of primary interest in the solar context and is worthy of further study.

\section{Acknowledgements}

LJS wishes to thank the International Astronomy Union and the Royal Astronomical Society for the award of travel grants that enabled her to participate in this meeting. NHB acknowledges NASA grant NNX07AL74G.  These computations were carried out at the UKMHD facility at St. Andrews University, supported by the UK STFC.

%\vfil\eject

\begin{discussion}

\discuss{Rogers}{The $R_I$ numbers isn't strictly large in the solar
  tachocline. So have you tried to run simulations in which $N$ is
  varying rapidly, as in the solar tachocline?}

\discuss{Silvers}{So far we have only looked at this one case which is mildly stratified.  It would indeed be interesting to look at cases with different, possibly more relevant, stratifications.}

\discuss{Hughes} {I believe that the time scale for stretching out the
magnetic field is important. Where does it appear in the criterion that
you displayed?}

\discuss{Silvers} {I feel that that is a question that is probably best answered by Nic or Geoff as it is from their earlier paper together.}

\discuss{Brummell} {The Richardson number criterion of Vasil \& Brummell is basically a bound derived by examining the maximum value of the destabilizing gradients that can be built before the the induction of $B_x$ by the shear halts.  It is therefore a criterion that equates values at a particular time, and therefore the timescale drops out explicitly.  This time is essentially the Alfven travel time on the original vertical field across (half of) the localized layer.  Further details can be found in  \cite{Vasil2}.}

\end{discussion}

\end{document}